\pdfoutput=1
\pdfoutput=1
\pdfoutput=1
\pdfoutput=1
\pdfoutput=1
 \documentclass[final,1p,times]{amsart}
\usepackage{graphicx}
\usepackage{amssymb}
\newcommand{\E}{\mathbb{E}}

\newcommand{\Z}{\mathbb{Z}}

\newcommand{\V}{\mathbb{V}}
\newcommand{\PP}{\mathbb{P}}

\begin{document}

\title{Is the Percolation Probability on $\mathbb{Z}^d$ with Long Range
Connections Monotone? }

\author{A. P. F. Atman} 
\address{Departamento de F\'{\i}sica e Matem\'atica  and Instituto Nacional de Ci\^encia e Tecnologia - Sistemas Complexos 
Centro Federal de Educa\c c\~ao Tecnol\'ogica de Minas Gerais,CEFET--MG, Av. Amazonas 7675, Belo Horizonte, MG, Brasil.}
\thanks{atman@dppg.cefetmg.br} 

\author{B. N. B. de Lima} 
\address{Departamento de Matem\'atica, Universidade Federal de Minas Gerais, UFMG, 
Av. Ant\^onio Carlos, 6627, Belo Horizonte, MG, Brasil.}
\thanks{bnblima@mat.ufmg.br}	

\author{M. Schnabel} 
\address{Fachhochschule Jena Fachbereich SciTec, Carl-Zeiss-Promenade 2 D-07745, Jena, Germany.}

\maketitle

\begin{abstract}

We present a numerical study for the threshold percolation probability, $p_c$, in the bond percolation model with 
multiple ranges, in the square lattice. A recent Theorem demonstrated by de Lima {\it et al.} [B. N. B. de Lima, R. P. Sanchis, 
R. W. C. Silva, STOCHASTIC PROC APPL {\bf 121}, 2043-2048 (2011)] states that the limit value of $p_c$ when the long ranges 
go to infinity converges to the bond percolation threshold in the hypercubic lattice, $\mathbb{Z}^d$, for some appropriate dimension $d$.
We present the first numerical estimations for the percolation threshold considering two-range and three-range versions of the model. 
Applying a finite size analysis to the simulation data, we sketch the dependence of $p_c$ in function of the range of the largest bond.
We shown that, for the two-range model, the percolation threshold is a non decreasing function, as conjectured in the cited work, and 
converges to the predicted value. However, the results to the three-range case exhibit a surprising non-monotonic behavior 
for specific combinations of the long range lengths, and the convergence to the predicted value is less evident, raising new 
questionings on this fascinating problem. 

\end{abstract}

%\begin{keyword}
%multiple range percolation \sep simulation \sep long range percolation threshold
%\end{keyword}

%\PACS 05.10.-a \sep 02.50.-r \sep 02.70.-c'

% main text
%\end{frontmatter}

\section{Introduction}

\label{intro}

Percolation is one of the most studied phenomena in statistical physics since it is closely related to several phase transitions in a 
wide range of systems. As a typical interdisciplinary subject, the literature shows contributions of researchers from different areas as 
mathematics, physics, civil engineering, soil science, hydrology etc. treating a wide range of problems from infiltration and porous 
flow to jamming and glassy transitions, passing by order-disorder models and magnetism \cite{stauffer94,G}.
A definition to percolation could be simply stated as following: considers a regular two-dimensional square lattice ($\Z \times \Z$ )
in each the vertex are connected by bonds which are present with probability $p$; with we start with one single bond occupied and 
randomly sort a new bond to connect two neighbors sites, eventually a large cluster will span along the entire system and we say that the 
system has percolate. 

Despite its simplicity, several open problems still challenge any theoretical treatment; for example, the exact  
value for the threshold percolation probability, $p_c$, for the regular lattice in three dimensions is not known, and even the continuity of 
the probability density function at $p_c$ still is an open problem ( for general facts in percolation see \cite{G}. The major importance in 
determine the percolation threshold 
and how it depends on the dimensionality of the space, $D$, is due to the link between percolation and phase transitions in natural 
systems. Several emergent phenomena arise near to the critical point, where power-law and fractal behaviors are observed. Most of these 
phenomena could be explained considering the cluster size distribution of similar components along the system, and percolation plays a major 
role to understand the mechanism involved in these natural processes. 

In the recent years, a particular attention was paid in literature on the study of irregular lattices, or complex networks. A central 
characteristic of these networks is an inhomogeneous distribution of bonds, or links, by node, and even a wide distribution of bond lengths. 
Hence, a natural question arises: what is the percolation threshold and how it behaves in function of the 
network parameters? Several works \cite{barab} have studied this problem, but most of them have focused the applications, {\it e. g.}, the 
Ising model in complex networks \cite{dorog}. Here, we present a more generic study of this problem from the percolation theory point of view; 
we introduce a percolation model with multiple bond ranges extending the usual bond percolation definition in the square lattice: 
we draw a bond to connect two sites in the lattice randomly chosen among a list of available sites within a given range. Thus, by means of 
this generalization of the standard model we expect to unveil possible hidden features which arise when we combine several 
bond ranges, and expect to shed some light in the problem of percolation in more complex networks.

Our primary aim is to verify a very recent conjecture from de Lima et al \cite{LSS} which states that the 
percolation threshold should be a decreasing function of the second range, when two bond ranges, $(1,k)$, are available, and 
Theorem 1 in \cite{LSS} which states that the $p_c$ value in this case converges to the bond percolation threshold in a $d=4$ 
hypercubic lattice in the limit of $k$ going to infinite. It worth to mention that add a new bond range is equivalent to increase 
the freedom degrees of the system, thus an operation analogous to increase the spatial dimension $D$.
Here, we present an extensive numerical study of the percolation threshold, $p_c$, for two-range, $(1,k)$, and three-range, $(1,k,m)$, 
bond percolation models. In the three-range model, the predicted value when $k \rightarrow \infty$ and $(m/k) \rightarrow \infty$, is
the percolation threshold of the $d=6$ hypercubic lattice. Surprisingly, the numerical results have shown a non-monotonic behavior for specific 
combinations of $k$ and $m$ which are not predicted by the theory.

The paper is structured as following: in order to well state the problem and present the analytical conjectures, we present the 
model definition and the theoretical predictions in the limit of diverging ranges in the next section. The numerical 
procedure and simulation results are discussed in the following and we drawn some conclusions and perspectives at the final section.

\section{Theoretical Aspects}
\label{theoric}

Let $G=(\V,\E)$ be a graph with a countably infinite vertex set $\V$. Consider the Bernoulli site percolation model on $G$ 
which associates to each vertex the values 1 (``occupied'') and 0 (``vacant'') with probability $p$ and $1-p$ respectively. 
This can be done considering the probability space $(\Omega, \mathcal{F},\PP_p)$, where $\Omega=\{0,1\}^{\V}$, 
$\mathcal{F}$ is the $\sigma$-algebra generated by the cylinder sets in $\Omega$ and $\PP_p=\prod_{v\in\V}\mu(v)$ is the 
product of Bernoulli measures with parameter $p$. We denote a typical element of $\Omega$ by $\omega$.

Given two vertices $v$ and $u$, we say that $v$ and $u$ are connected in the configuration $\omega$ if there exists a finite 
path $\langle v=v_0,v_1,\dots,v_n=u\rangle$ of occupied vertices in $\V$, such that $v_i\neq v_j,\ \forall i\neq j$ and 
$\langle v_i,v_{i+1} \rangle$ belongs to $\E$ for all $i=0,1,\dots,n-1$. We will use the short notation 
$\{v\leftrightarrow u\}$ to denote the set of configurations where $u$ and $v$ are connected.

For a given vertex $v$, the cluster of $v$ in the configuration 
$\omega$ is the set $C_v(\omega)=\{u\in \V; v\leftrightarrow u \mbox{ on } \omega\}$. We say that the vertex $v$
percolates when the cardinality of $C_v(\omega)$ is infinite; we will use the following standard notation
$\{v\leftrightarrow\infty\}\equiv\{\omega\in \Omega; \#C_v(\omega)=\infty\}$. 
Once $v$ fixed, we define the percolation probability of the vertex $v$ as the function 
$\theta_v(p):[0,1]\mapsto [0,1]$ with $\theta_v(p)=\PP_p(v\leftrightarrow\infty)$. The percolation threshold 
(or critical point), $p_c(G)$, is defined by $$p_c(G)=\sup\{p\in[0,1]; \theta(p)=0\}~.$$

From now on, the vertex set $\V$ will be $\Z^d,\ d\geq 2$ and for each positive integer $k$ define
\begin{eqnarray*}
\E_k & = & \{\langle (v_1,\dots,v_d),(u_1,\dots, u_d)\rangle\in \V\times\V; \exists ! i\in\{1,\dots,d\} \\
     &   &  \textrm{such that }|v_i-u_i|=k ~\textrm{and} ~v_j= u_j,~\forall~~ j\neq i\}~.
\end{eqnarray*}
Let's define the graph $G^k=(\V, \E_1\cup\E_k)$ that is, $G^k$ is the $\Z^d$ equipped with nearest neighbors 
bonds and long range bonds with length $k$ parallel to some coordinate axis. Observe that $G^k$ is a transitive graph, 
hence the function $\theta_v(p)^k$ does not depend on $v$ and we write only $\theta^k(p)$ to denote
$\PP_p(0\leftrightarrow\infty)$ on the graph $G^k$.

Consider the sequence, $(p_c(G^k))_k$, of the percolation thresholds of the graphs $G^k$. In \cite{LSS} (see Theorem 1), 
it is proven that $$\lim_{k\rightarrow +\infty} p_c(G^k) = p_c(\Z^{2d}),\forall d\geq 2~,$$
that is, the percolation threshold of $G^k$ converges to the percolation threshold of the hypercubic (nearest neighbor) 
lattice in $2d$ dimensions. It is also conjectured \cite{LSS} that the sequence $(p_c(G^k))_k$ is non increasing in 
$k$. One goal of this work is to simulate long range percolation on the graph $G_k$ in $d=2$. These simulations, as 
described in the next section, show that $p_c(G^k)$ goes to $p_c(\Z^4)$ monotonically, confirming the Theorem 1 and the 
Conjecture in \cite{LSS}.

Theorem 1 and the Conjecture in \cite{LSS} can be generalized for bond percolation with multiple ranges. Given a sequence
$\vec{k}=(k_1,\dots,k_n)$ with $k_i\in \{2,3,\dots\},\forall i$,  consider now the graph $G^{\overrightarrow{k}}$ as
$(\Z^d,\E_1\cup(\cup_{i=1}^n \E_{k_1\times\dots\times k_i}))$. That is,  $G^{\overrightarrow{k}}$ is $\Z^d$ 
decorated with all bonds parallel to each coordinate axis with lengths 
$1,k_1,k_1\times k_2,\dots, k_1\times k_2\times\dots\times k_n.$ In this case Theorem 1 of \cite{LSS} states that
$$\label{eqprin}\lim_{k_i\rightarrow\infty,\forall i}p_c(G^{\vec{k}})=p_c(\Z^{d(n+1)}),\forall d\geq 2$$ and the 
Conjecture says that the probability of percolation on the graph $G^{\overrightarrow{k}}$ is non decreasing in 
each variable $k_i$.

In this work, we also perform simulations in the case $n=3$, considering bonds with three different ranges 1, $k$ and $m$. 
The results support the Theorem 1 and the Conjecture in \cite{LSS} (in this case, $m$ should be a multiple of $k$). One 
interesting feature displayed by simulations is that the percolation threshold is not monotone in $k$, with $m$ fixed.

\begin{figure}[t]
\begin{center}
\includegraphics[width=0.8\linewidth]{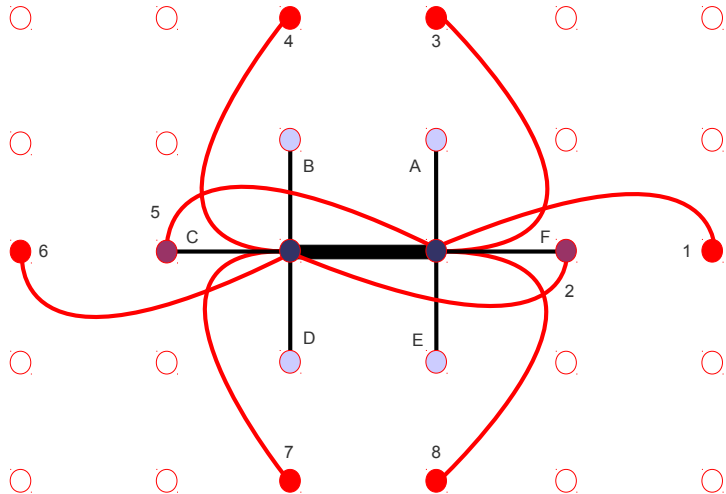}
\end{center}
\caption{Neighborhood of a bond in the 2-range version of the model. The first neighbors of the central black bond 
are shown, for $(1,k) = (1,2)$. Note that there is 6 short-range bonds (A-F) with the same length as the central bond, 
and 8 long-range bonds (1-8). The sites connected to the central bond are filled with different gray levels, depending 
on the range of the bond linked to the site.
\label{fig:fig01}}
\end{figure}

\section{Numerical Procedure}
\label{numerical}
\begin{figure}[t]
\includegraphics[width=\linewidth,angle=0]{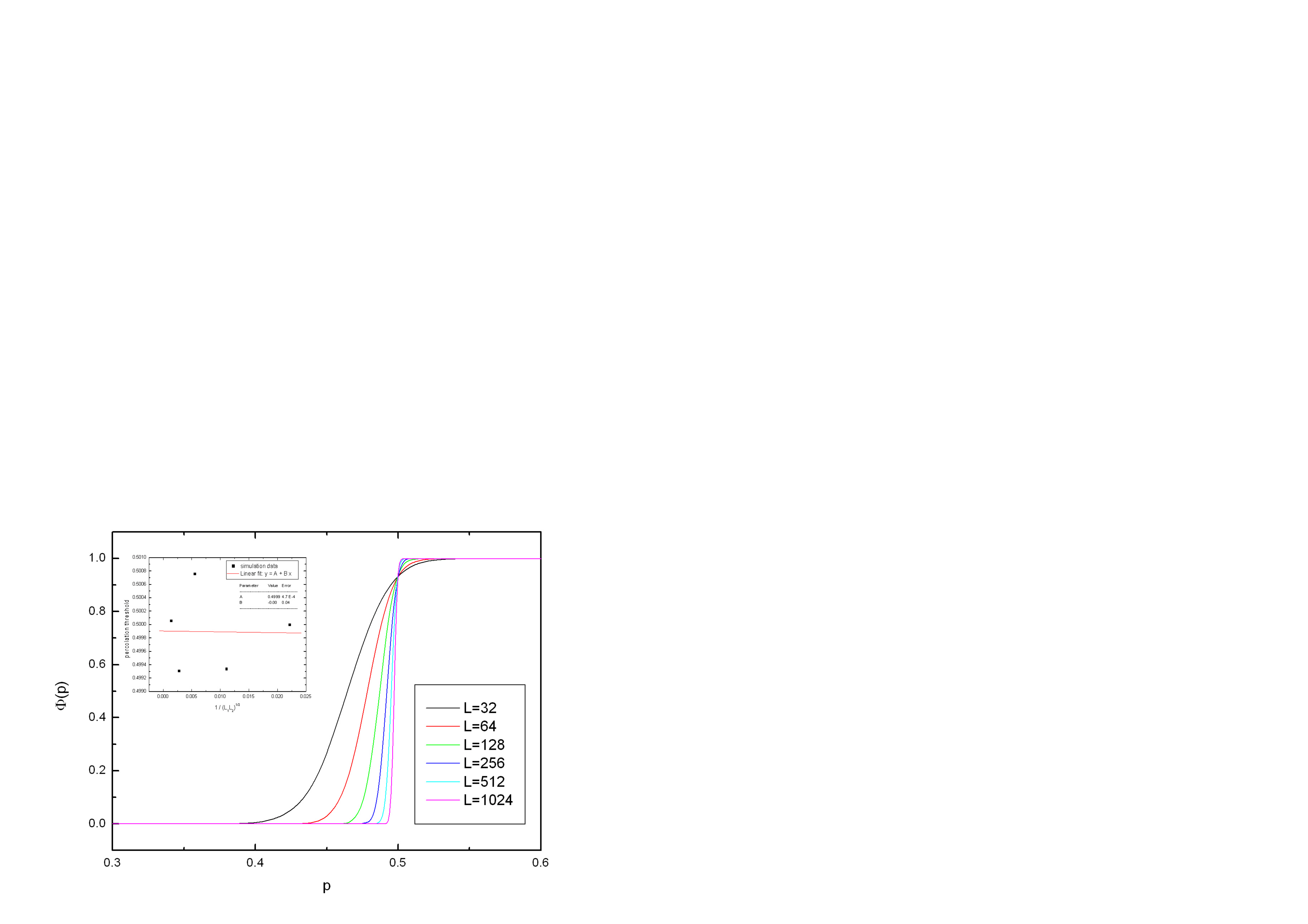}
\caption{Probability of a site belongs to the infinite cluster in a single range bond percolation model. 
The plot shows the variation of $\Phi(p)$ to 6 values of $L=32,64,128,256,512,1024$. Note that all curves tend
to cross at the same point. Considering two consecutive system sizes, we measure the ordinate value of the crossing
point. The inset shows the crossing values obtained in function of the inverse of the geometric mean between the 
system sizes, $1/ \sqrt{L_i L_j}$. The independent term of the linear fit correspond to the best estimative for
the percolation threshold to a infinite system size, and matches perfectly with the exact value $p_c=1/2$.
\label{fig:bond1r}}
\end{figure}

In this section we present the numerical approach used in the simulations. We consider a square two-dimensional regular 
lattice as the site substrate of the model. When a link to a given range is open, two sites in a column, or in a line, 
will be connected to each other forming a bond - see Figure \ref{fig:fig01}. For each bond, besides to the six neighbors
in the same range, there are eight neighbors consisting of bonds of another length. Thus, in the 2-range version
of the model there are 14 neighbors and, in the 3-range, 22 neighbors to be taken into account for every new bond opened.

The algorithm used to estimate the percolation threshold is based on the work of Newman and Ziff \cite{ziff01}. 
Basically, we label each bond of the lattice assigning the first $2N$ natural numbers to bonds of unitary length 
(where $N=L \times L$, and $L$ is the system size); the labels $2N+1$ to $4N$ are assigned to the range $k$, and so on. 
Thus, we enumerate all the possible bonds to be opened in the system from $0$ to $r~2N$, where $r$ is the number of ranges 
considered.

\begin{figure}[t]
\includegraphics[width=0.95\linewidth]{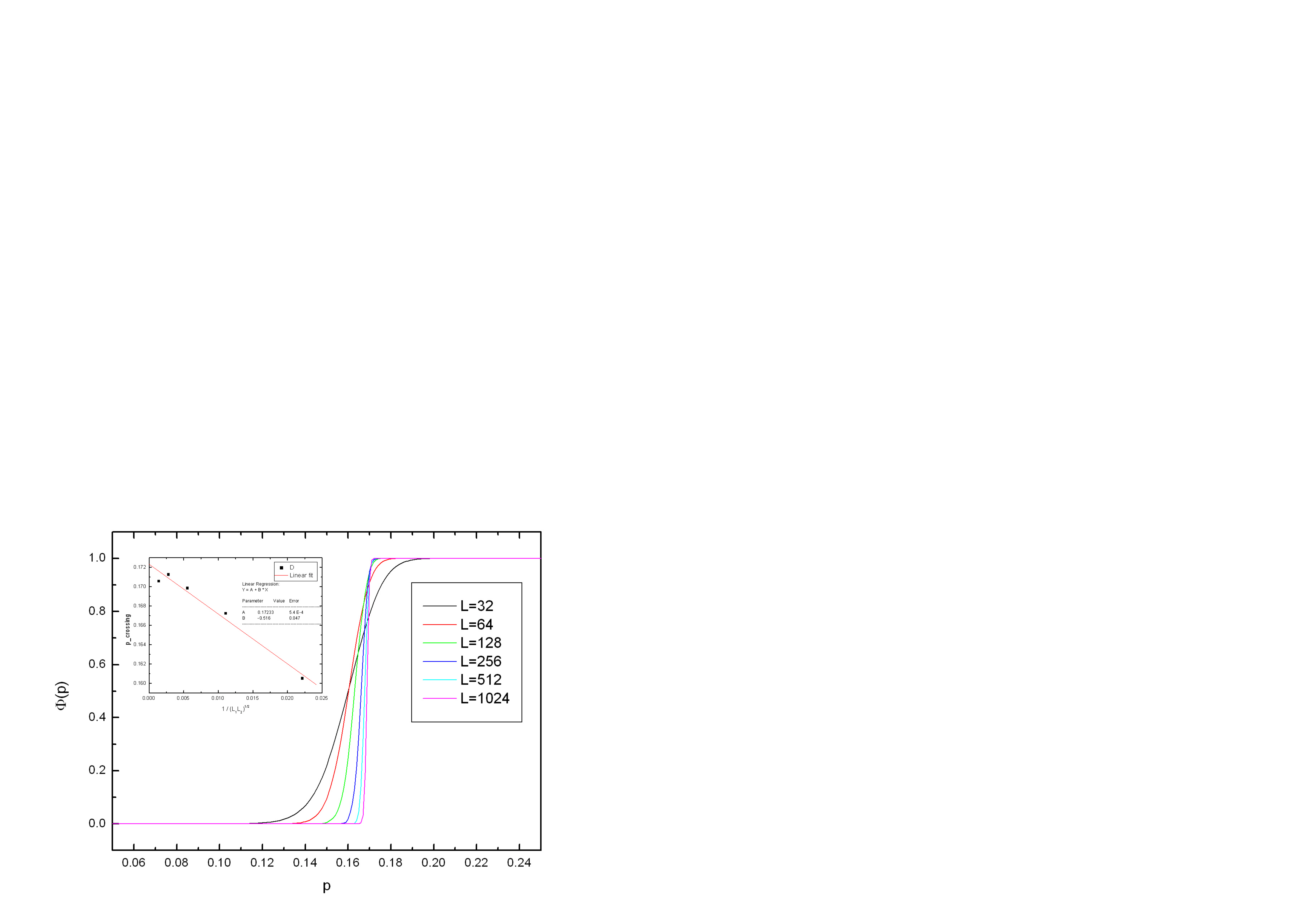}
\caption{Probability of a site to belong to the infinite cluster in a 2-range $(1,k)$ bond percolation model, with $k=7$. 
The plot shows the value of $\Phi(p)$ considering 6 values of $L=32,64,128,256,512, 1024$. The inset shows the 
finite size analysis used to obtain the best estimative for the percolation threshold at the infinite system (see the 
caption of Figure \ref{fig:bond1r}). This procedure is repeated for each value of $k$ considered in this version 
of the model, and the dependence of the threshold percolation in function of $k$ is shown in Figure \ref{fig:pc2r} .
\label{fig:bond2r}}
\end{figure}

In this way, each realization consists of a random permutation in the list of the bonds to be opened, varying the occupation 
probability, $p$, from 0 to 1. After each bond is open, the percolation condition is tested using the modified 
Hoshen-Kopelman algorithm to identify the clusters \cite{stauffer94}. The probability of a given site to belong to the 
infinite cluster is then estimated from the simulated data using the normalization procedure described in \cite{ziff01}.

The code was developed in C and Purebasic $\textregistered$ languages. Purebasic $\textregistered$ code developed by M. Schnabel 
is procedure-oriented and 
1000 samples of a system with $L=1024$, single range percolation model, takes around two minutes to run in an AMD Athlon X64 
1.9 GHz and 3 GBytes DDR2 RAM machine. It is about $20 \%$ faster then the similar C implementation.

In order to validate the code developed, we initially consider the bond percolation model limited to a single range,
in a square lattice of linear extension $L$, with periodic boundary conditions in both directions. We simulate
1000 realizations for each system size, $16 < L < 4096$, and calculate the probability of a given site belongs to 
the infinite cluster, $\Phi(p)$, Figure \ref{fig:bond1r}. In Figure 
\ref{fig:bond2r} we show the corresponding calculation for the two-range bond model considering $k=7$. We denote by
$(1,k)$ or $(1,k,m)$ the 2-bond and 3-bond models, respectively, where $k, m \in N$.

The percolation criterion used assumes that the infinite cluster arises when there is a path of connected bonds crossing the 
entire system, vertically and horizontally, not using the periodic boundaries. Although the particular choice 
used for this criterion, we believe that we would get qualitatively identical results even for other choices, as 
discussed in the next section

\section{Results and Discussion}
\label{results}

\begin{figure}[t]
\includegraphics[clip,width=\linewidth,clip,angle=0]{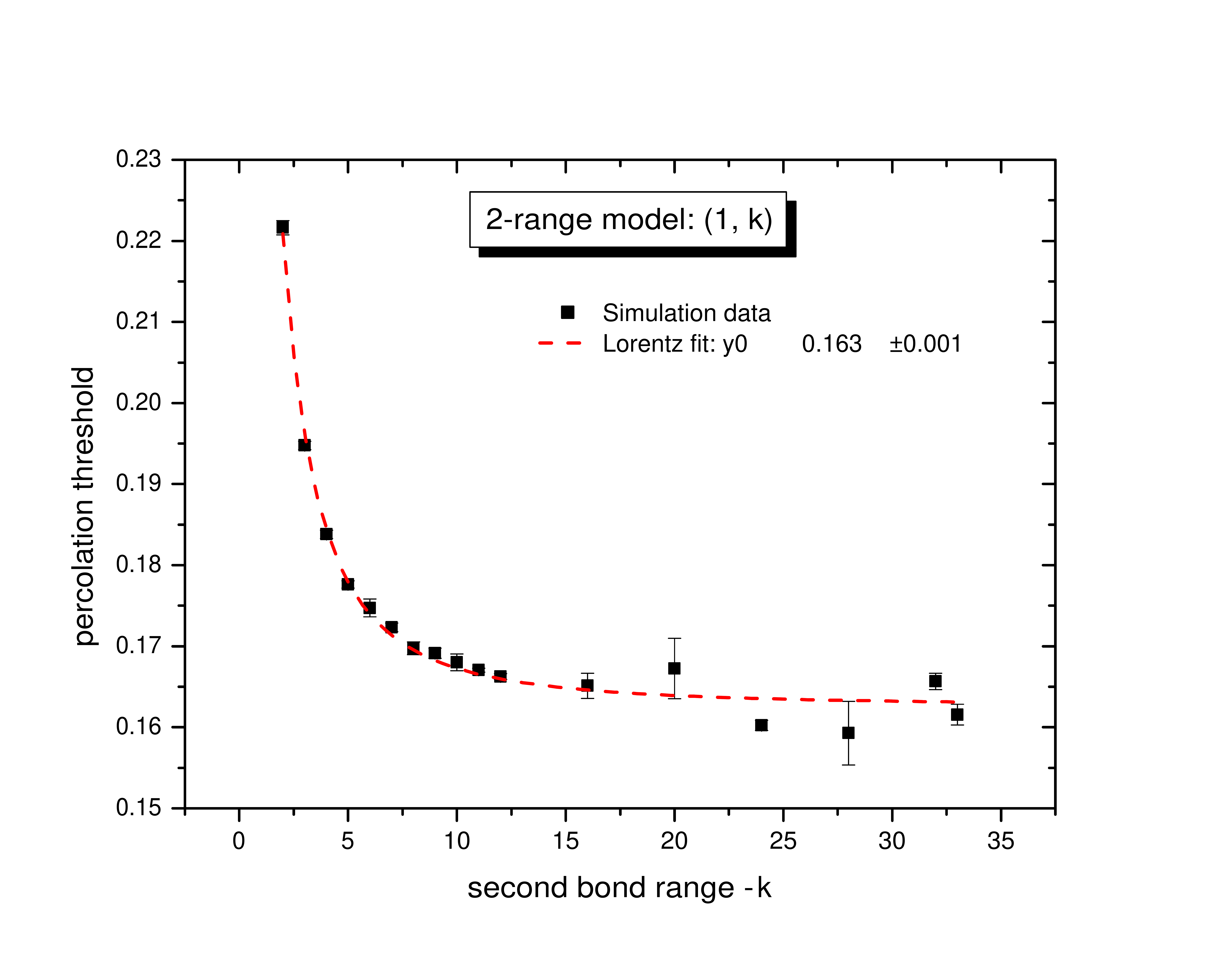}
\caption{Threshold percolation probability, $p_c$, for 2-range bond models, with increasing range $k$. There is no 
justification for the fit function used and the line is only to guide the eyes. The limit value of $p_c$ is obtained
from the fitting parameter $y0$. 
\label{fig:pc2r}}
\end{figure}

Figure \ref{fig:bond2r} presents the procedure used estimate the percolation probability to the 2-range 
model $(1,k)$, with $k=7$. Considering different values of $k$, it is straightforward the calculation of $p_c^k$ for 
increasing $k$, which is shown in Figure \ref{fig:pc2r}. The limiting value for $k \rightarrow \infty$ is 
$p_c \sim 0.162(2)$, close to the expected value for the bond percolation threshold at the $d=4$ hypercubic lattice - 
$p_c = 0.160131(1)$ \cite{grassberger03} as stated by Theorem 1 \cite{LSS}. Thus, we can conclude from this plot that 
the percolation threshold exhibits a monotonic dependence on $k$, supporting the Conjecture in \cite{LSS}, 
and confirming the prediction for the limit $k \rightarrow \infty$ stated by Theorem 1.

\begin{figure}[t]
\includegraphics[width=\linewidth,clip,angle=0]{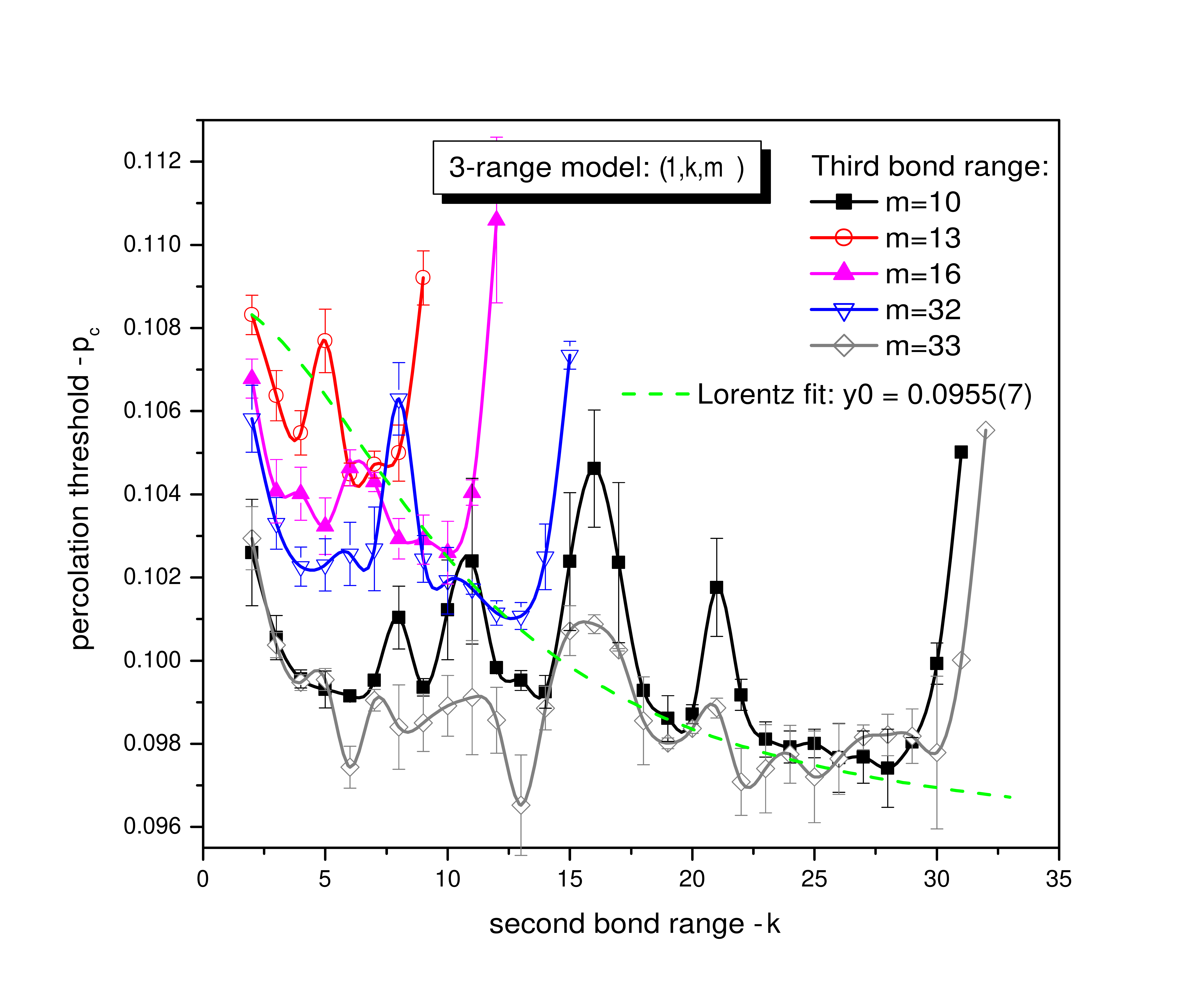}
\caption{Threshold percolation probability, $p_c$, for 3-range bond models, in function of the range $k$; five different 
curves corresponding to increasing $m$ are shown. Note the differences when $m$ is odd or a prime number, smoothing the 
curves. The striking non monotonic behavior observed for specific combinations of the long-ranges still is not fully 
understood to the moment. The dashed line represents a Lorentzian fit to the minima of the plots.
\label{fig:pc3r}}
\end{figure}

The results for the 3-range model are shown in Figure \ref{fig:pc3r}. We consider five different values to $m$, and all the 
possible $k$ between 1 and $m$. All results are shown in Figure \ref{fig:pc3r}. We can observe some striking features: 
first, there is a clear downward trend of $p_c$ with increasing $m$. To $m$ fixed, generally $p_c$ decreases with 
$k$, except when $k$ approaches to $m$, or to the half-value of $m$. This trend was observed for $m$ even, odd or prime. 
For $m=32$ a curious feature appeared: the value of $p_c$ increases about $k=11$ and $k=21$. We do not have a good 
explanation for this match so far. The increase to $k=8$ can be also be attributed to the correspondence 
with $m=32$, as 32 is multiple of 8; however, for ranges of $k=4$ and $m=16$, for example, a similar effect was not
observed, discarding this conjecture.

Considering the curves for $m=10,13$ and $16$, they all behave similarly, with $p_c$ decreasing with $k$ except for $k$ around   
to $m$ or $m/2$. For $m=13$, the increasing at $m/2$ is more discreet, but occurs between $k=6$ and $7$, since the 
half-value of $m$ is not an integer. The curves for $m=32$ and $33$ show different features. We observe that the decrease 
of $p_c$ with $k$ is more subtle, but the increasing at $m$ and $m/2$ is quite evident. The peaks observed at $k=8,11,21$ are 
unexpected and not well understood for the moment. The limit value of $p_c \simeq 0.0955(7)$ around to $k=25$, approaches 
to the bond threshold percolation value at $d=6$ hypercubic lattice, $p_c=0.0942019(6)$ \cite{grassberger03}, which is expected 
when the two long ranges tend to infinity.

\section{Conclusions and Perspectives}

We present by the first time the percolation threshold estimation for a multiple range bond percolation model. We study the 
behavior of $p_c$ in two and three range models in function of the bond range using numerical simulations. We have confirmed the 
predicted value to the percolation threshold when the long ranges tend to infinity in these models - Theorem 1 in \cite{LSS}. 
A remarkable effect was observed for the 3-range version of the model, with an unexpected increase in the percolation threshold
for some values specific values of the second and third ranges which defy any convincing explanation to the present date and 
could represent an important feature in models with long range interactions or even systems dealing to complex networks.

\noindent {\bf Acknowledgements} The authors are indebted to Lucas Alves Martins by initial work on the codes, and to 
N. Moloney by the kindly cession of percolation codes to validate our initial algorithm.

APFA is partially supported by CNPq and FAPEMIG, BNBL is partially supported by CNPq and MS would like to thank IASTE and 
CEFET-MG.


\begin{thebibliography}{00}

\bibitem{stauffer94}
D.Stauffer, A. Aharony, {\it Introduction To Percolation Theory}, 2 ed.,
(Taylor $\&$ Francis, London, 1994)

\bibitem{G}
Grimmett G., \emph{Percolation}, 2nd edition, Springer-Verlag, Berlin, (1999).

\bibitem{LSS}
B. N. B. de Lima, R. P. Sanchis, R. W. C. Silva, {\it Critical Point and Percolation Probability in a Long Range 
Percolation Model on $\mathbb{Z}^d$.} Stochastic Processes and their Applications {\bf 121}, 2043-2048 (2011).

\bibitem{ziff00}
M. E. J. Newman and R. M. Ziff, {\it Efficient Monte Carlo algorithm and high-precision results for percolation},  
Physical Review Letters {\bf 85} (19): 4104-4107 (2000).

\bibitem{barab}
A.-L. Barab\'asi, R. Albert, H. Jeong
\newblock{Physica A: Statistical Mechanics and its Applications {\bf 272}, 173–187 (1999).}

\bibitem{dorog}
S. N. Dorogovtsev, A. V. Goltsev, J. F. F. Mendes
\newblock{Review of Modern Physics {\bf 80}, 1275–1335 (2008).}

\bibitem{ziff01}
M. E. J. Newman and R. M. Ziff, {\it A fast Monte Carlo algorithm for site or bond percolation}, Physical Review E {\bf 64}, 
016706 (2001).

\bibitem{grassberger03}
Grassberger, P. {\it Critical percolation in high dimensions} Phys. Rev. E {\bf 67} (3): 4 (2003).

\end{thebibliography}
\end{document}